\documentclass[conference]{IEEEtran}
\IEEEoverridecommandlockouts
\usepackage{cite}
\usepackage{amsmath,amssymb,amsfonts}
\usepackage{cleveref}
\usepackage{algorithmic}
\usepackage{graphicx}
\usepackage{textcomp}
\usepackage{multirow}
\usepackage{comment}
\usepackage{xcolor}
\usepackage[font={small}]{caption}
\usepackage{tabularx}
\usepackage{subcaption, mwe}
\usepackage{placeins}
\usepackage{float}

\def\BibTeX{{\rm B\kern-.05em{\sc i\kern-.025em b}\kern-.08em
    T\kern-.1667em\lower.7ex\hbox{E}\kern-.125emX}}
  \usepackage[multiple]{footmisc}
  \newcolumntype{C}[1]{>{\centering\let\newline\\\arraybackslash\hspace{0pt}}m{#1}}
\begin{document}

\title{Crime Rate Prediction with Region Risk and Movement Patterns}

\author{\IEEEauthorblockN{Shakila Khan Rumi}
\IEEEauthorblockA{
\textit{RMIT University}\\
Melbourne, VIC \\
shakilakhan.rumi@rmit.edu.au}
\and
\IEEEauthorblockN{Flora D. Salim}
\IEEEauthorblockA{\textit{RMIT University}\\
Melbourne, VIC \\
flora.salim@rmit.edu.au}

}

\maketitle

\begin{abstract}
The location-based social network, \textit{Foursquare}, reflects the human activities of a city. The mobility dynamics inferred from \textit{Foursquare} helps us understanding urban social events like crime In this paper, we propose a directed graph from the aggregated movement between regions using \textit{Foursquare} data. We derive \textit{region risk factor} from the movement direction, quantity and crime history in different periods of the day. Later, we propose a new set of features,  DIrected graph Flow FEatuRes (DIFFER) which are associated with \textit{region risk factor}. The reliable correlations between DIFFER and crime count are observed. We verify the effectiveness of the DIFFER in monthly crime count using Linear, XGBoost, and Random Forest regression in two cities, Chicago and New York City.
\end{abstract}


\section{Introduction}
Finding ways to control a city's crime rate is important to ensure safe and secure living space in a society. According to criminology theory, the surrounding environment, including neighbourhood regions, and the movement of people, play a crucial role in crime event prediction~~\cite{cohen1979social, brantingham1993environment,miller2005measurement}. The widespread use of location-based social networks such as \textit{Foursquare} opens a door of opportunities to analyse the crime events based on human movements. Understanding human movements in a city can provide better recommendation in city planning~\cite{sadri2018will}. In this paper, we study crime rate prediction with the help of urban mobility data. 

\begin{figure}[!h]
	\centering
	\begin{subfigure}[t]{.23\textwidth}
		\centering
		\includegraphics[width=\textwidth]{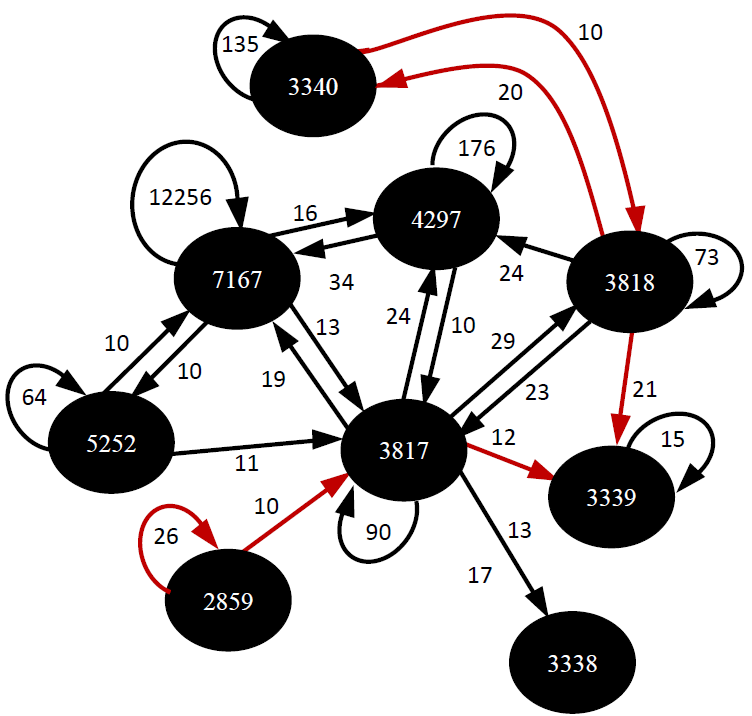}
		\caption{Morning}
		\label{fig:1}
	\end{subfigure}
	\centering
	\begin{subfigure}[t]{.23\textwidth}
		\centering
		\includegraphics[width=\textwidth]{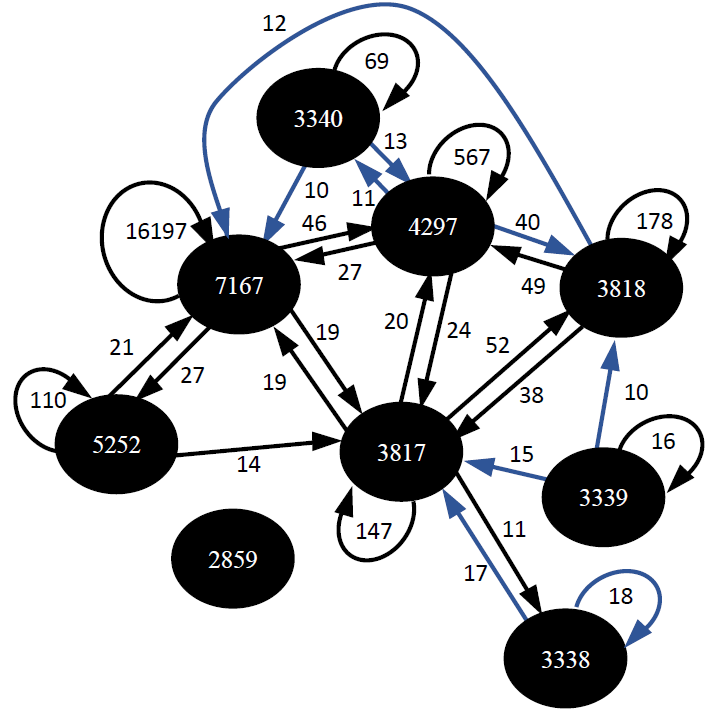}
		\caption{Afternoon}
			\label{fig:2}
	\end{subfigure}
	\caption{Check-in movements for a few venues in New York in the mornings and afternoons of March, 2018}
	\label{fig:All}
\end{figure}

Recently, there has been research linking crime events with urban dynamics using \textit{Foursquare} data~\cite{rumi2018theft}. However, this study focused on a region’s check-in information to predict crime events. There has been no focus on linking human movement between two regions. The movement between regions using taxi flow data was considered for crime inference in \cite{wang2016crime}, however, did not account for the variation of movement in different time periods of day. For example, people who move from their home to work will move in the opposite direction later in afternoon. In Figure~\ref{fig:All}, we observe that considering \textit{Foursquare} movement data, the direction and number of people moving in the morning and afternoon for New York City are different. The red edges between two regions in morning denote the links, if they are different than afternoon movement direction. On opposite, the blue edges between two regions in afternoon denote the links, if they are different than morning movement direction. Based on this finding, in this paper, we analyse the association between crime rate and human mobility in different periods of the day.  

We propose a directed weighted graph based on the movement in \textit{Foursquare} check-ins. Graph based solution was applied in past for contextual recommendation~\cite{ren2017location}, however in this study our aim is to measure graph flow based features for crime prediction. We introduce a new value to graph node, \textit{region risk factor}, which is derived based on mass human movement between regions over a certain period of the day. The hypothesis is that human mobility from a high crime risk area implies a high crime risk in the arrival area. Suppose, a region has 200 arrival check-ins in a certain period. If many of the check-ins are from regions with high crime rates, the check-ins pose a high risk of crime events happening in the in-flowing region at that time interval. It is to be noted that the people who are moving from high crime areas are not associated with crime event occurrences. We propose a set of graph flow based features, DIFFER from \textit{region risk factor} of the connected regions. These features represent hyperlinks between regions in different periods of a day by mass human movement.

The contributions of this paper are summarized as follows:
\begin{itemize}
    \item This is the first work that predicts the crime rate based on the hyperlinks between regions during different periods of a day. A new component, \textit{region risk factor} is proposed to combine the regional crime risk with human movement.
    \item A set of new features, DIFFER, from \textit{region risk factor} and directed mobility graph is crafted. 
    \item This work verifies the effectiveness of DIFFER in the crime rate prediction problem using regression analysis and significance test. Real-world crime data and \textit{Foursquare} movement data are used for evaluation. The experimental results show that the proposed features are highly correlated with the crime count of a region. 
\end{itemize} 

\section{Related Work}
In previous studies, data from multiple sources including historical, geographic, demographic and social media were used as input variables in the crime prediction task. In~\cite{UCIcrime}, the authors derive multiple features based on census and crime data. In~\cite{kearns2017preventing}, the authors predicted whether the crime rate would be above 70\% using the extracted features. There were also a few previous data mining studies, which were devoted to verify the impact of human mobility onto crime. In~\cite{Bogomolov2014,bogomolov2015moves}, the authors extracted human behavior from mobile network activities and demographic features of people connected to the network over different regions. The study showed that the
combination of mobile activity data and demographic data can be used to predict crime events with better accuracy. \textit{Foursquare} check-in data was used to measure the ambient population of a region and to understand the long-term crime event occurrences~\cite{kadar2016exploring,Kadar2018}. In~\cite{rumi2018,rumi2020realtime} the authors measured proposed several dynamic features using \textit{Foursquare} data to measure the social diversity of a region, and predict short-term crime event occurrences. In~\cite{rumi2018theft}, the authors proposed crime-specific dynamic features by analyzing the individual risk factor of the users and extracted multiple features based on the risk analysis.

However, these works did not explore the correlation of crime events occurrences in a place with the mobility flow in that region. Thus, these studies could not capture the social interaction between two regions in crime event prediction task. The mobility data represented by taxi flows and Points of Interest (POIs) can improve crime rate inference~\cite{wang2016crime}. Here, the authors’ hypothesis is that the social interaction between two places can be inferred through taxi trips, and crime propagates based on these connections.  

To the best of our knowledge, none of the previous works has correlated the large scale human movement in between regions during different time period of the day with the crime rate in a region. Our work attempts to fill this gap. 
\section{Dataset Description}\label{Dataset}
 The datasets are collected for Chicago and New York City. We collect different types of data including check-ins and crime events for each city. We segment each city into a $400\times 400 $ grid.
 
 \subsection{POI and Check-in Data}
 The datasets are collected for Chicago and New York City. We collect POI, check-ins and crime events for each city. We segment each city into a $400 \times 400 $ grid.\\
\textbf{POI and Check-in Data}
The POI and check-in information is collected from \textit{Foursquare}. This dataset is provided as part of the \textit{Future Cities Challenge} at Netmob~\footnote{https://www.futurecitieschallenge.com/}. The check-in information describes an aggregated count of all movements from one venue to another, separated by month and five time intervals: \textit{Morning, Midday, Afternoon, Night,} and \textit{Overnight}.\\
\textbf{Crime Data}
We collect the 2018 crime event records for Chicago and New York from the Open Data Portals of the respective city councils~\footnote{https://data.cityofchicago.org/Public-Safety/Crimes-2001-to-present/ijzp-q8t2}\footnote{https://data.cityofnewyork.us/Public-Safety/NYC-crime/qb7u-rbmr}. Each dataset consists of the longitude, latitude, and the time and date of crime event occurrences. The total number of crime event occurrences are 263,515 and 452,958 for Chicago and New York respectively.
\section{Mobility Associated Crime Risk Analysis}
Region associated crime risk factor analyses regional crime risk levels with the aggregated number of check-ins in that region. It captures the following intuition. If the incoming check-ins of a region are from a high crime rate area, it means that the risk of crime event occurrence is high in that region during that time interval. 
\subsection{Region Risk factor}
We analyse the risk associated with each node based on above-described intuition, named as \textit{region risk factor}. We hypothesise that the aggregated number of check-ins in a region grouped by its ancestor location is potential to disclose the crime risk level of that region. To formulate the \textit{region risk factor}, we consider a city as a directed weighted graph, $G = (V,E)$. Here, $V$ is the set of nodes on the graph which represents the $400 \times 400$ grids, and $E$ represents the set of edges between two grids based on the mobility direction of Users. In Figure~\ref{fig:All}, a part of New York City is represented as a  graph. Each node of the graph denotes $400 \times 400$ grids. The weight of the edge between nodes is determined using the check-in count in \textit{Foursquare} and the direction depends on check-in direction. Based on this graph, we calculate the \textit{region risk factor} of a node, $v \in V$, for time interval, $t$, is:
\begin{equation}
    RR(v,t)=\dfrac{|Cr(v,t)|}{|C(v,t)|},
\end{equation}
where $Cr(v,t)$ denotes the aggregated crime events that happened in node, $v$, in time interval, $t$ and $|C(v,t)|$ denotes the total number of check-ins. 
\subsection{Feature Description}
For each node $v \in V$  in a time interval, $t$, we derive different nodal and edge features from historical crime, POI and movement Data. Nodal features describe the characteristics of the focal grid only while edge features determine how the crime rate of a region is influenced by its connected region. 
\subsubsection{DIrected graph Flow FEatuRes (DIFFER)}
Several features are derived from \textit{region risk factor} for each grid in each time interval including \textit{risk distribution}, \textit{risk count}, \textit{risk ratio} and \textit{self risk}. Here, \textit{risk distribution}, \textit{risk count}, and \textit{risk ratio} are the edge features, which depend on the region risk factors of the connected regions. \textit{Self risk} represent the nodal feature. The description of the features are as follows:

\textbf{Risk Distribution}
The risk distribution consists of the mean and median of the \textit{region risk factor} associated with the regions $r\in R (v,t)$ from where people are moving to focal node, $v$, in time interval $t$.

\textbf{Risk Count}
The risk count in node, $v$, in time interval, $t$, determines the number of regions with high risk than average from where incoming movement occur in node, $v$, in that time interval. The risk count is denoted by,
\begin{equation}
    RC(v,t)=| \{r: r\in R(v,t) \textnormal{ and } RR(r,t)>\dfrac{1}{|R|}\sum_{n\in V}RR(n,t)\}|.
\end{equation}
Here,  $R(v,t)$ denotes the regions which are origin of check ins to node $v$, and $|R|$ is the total number of regions.

\textbf{Risk Ratio}
Risk count determines the absolute number of regions with high risk. We normalise this feature based on total regions with incoming movement. The risk ratio is modelled using,
\begin{equation}
    RT(v,t)=\dfrac{RC(v,t)}{|R(v,t)|}.
\end{equation}

\textbf{Self Risk}
This feature represents absolute value of region risk that is associated to the focal node, $v$ in time interval, $t$, $RR(v,t)$

\subsection{Crime History, POI and Mobility based Features}
For each region in a time interval, we further extract the historical and geographic features from crime history and POI data. In particular, we extract Crime Event Density and Neighbourhood Crime as historical feature and POI density, Venue Category Distribution and Venue Entropy as geographic features. We also extract several nodal features from the aggregated movements between regions including Incoming Movement, Outgoing Movement, Stationary Movement and Diversity of Movement. These features represent human dynamics in a region in a certain period of the day. If a check-in is performed from other nodes to node, v in a particular time interval, then it is considered as incoming movement for node, v for that time. On opposite, if a check-in is performed from node, v to other nodes, it is considered as outgoing movement for node, v. When the origin and destination of a check-in is the same node, it is denoted as stationary movement.  

\section{Model Description}
The main purpose of this paper is to show the effectiveness of the features derived from the directed movement graph in crime rate prediction. To do so, we apply Linear Regression (LR), XGBoost (XGB), and Random Forests (RF) regression to count the number of crime events of a node in a time interval. We compare the results from different regression algorithms to determine which model is the best to predict crime count. LR is a statistical model, in which, the response variable linearly depends on the explanatory variable~\cite{kutner2005applied}. In the model, each explanatory variable is associated with a non-random parameter called a dummy variable to represent the linear dependency. The model is derived based on the Gaussian or normal probability distribution function. RF is an ensemble learning based regression model~\cite{breiman2001statistical}. It constructs several trees based on random predictors and combines them with an ensemble learning method to predict the output variable. XGB is another tree-based ensemble learning regression model, which use gradient boosting framework for prediction~\cite{chen2016xgboost}.

In this study, we only use the region and time intervals where check-in movement data exists to train the model. For example, in a month, $m$, for a region, $v$, and over time interval, $t$, if there is any type of movement data point, it generates the training and test data based on the features described in previous section. Finally, the regression model is trained with different feature settings to know each feature's effectiveness. The performance metric compares which feature sets are significant for crime rate prediction promptly.

\def\arraystretch{1.4}
\section{Experiments}
\subsection{Settings}
The dataset used in this experiment is introduced in Section~\ref{Dataset}. Each day is segmented in five intervals and for each interval the are aggregated in monthly level. To prevent extreme sparsity situations, we only include check-in movement data with ten or more unique movements in a month. We partition the data points between January 2018 and September 2018 (inclusive) as the training data and the data points between October 2018 to December 2018 (inclusive) as the test data. If new regions are found with check-in movement greater than, or equal to 10 in test data for a time interval, the risk value for that region is set 0.

Two performance metrics, Mean Absolute Error (MAE) and Root Mean Squared Error (RMSE), are used to verify the effectiveness of the crime inference model.

\begin{figure}[!ht]
	\centering
	\begin{subfigure}[t]{.23\textwidth}
		\centering
		\includegraphics[width=\textwidth]{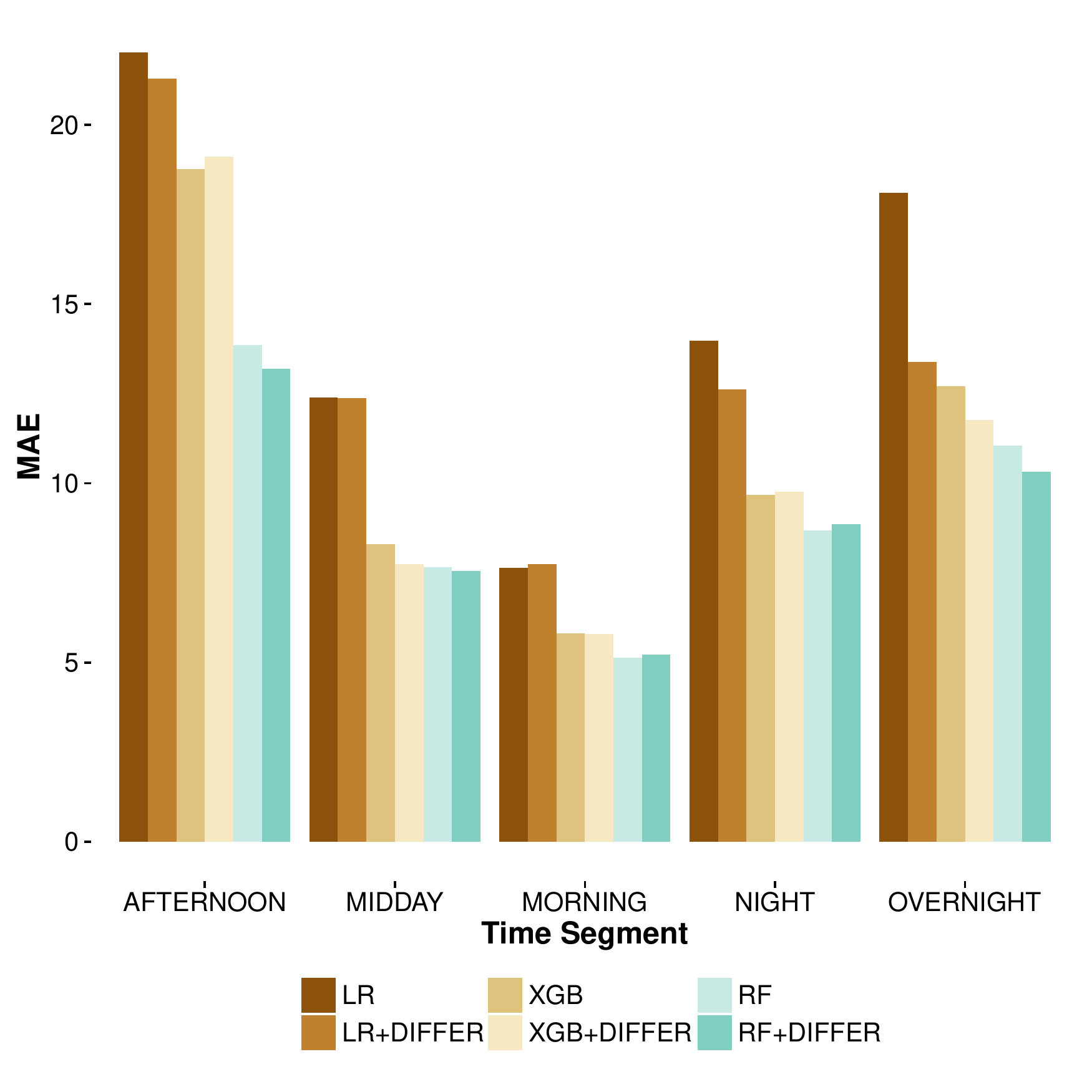}
		\label{fig:1_mae}
		\caption{New York}
	\end{subfigure}
	\centering
	\begin{subfigure}[t]{.23\textwidth}
		\centering
	\includegraphics[width=\textwidth]{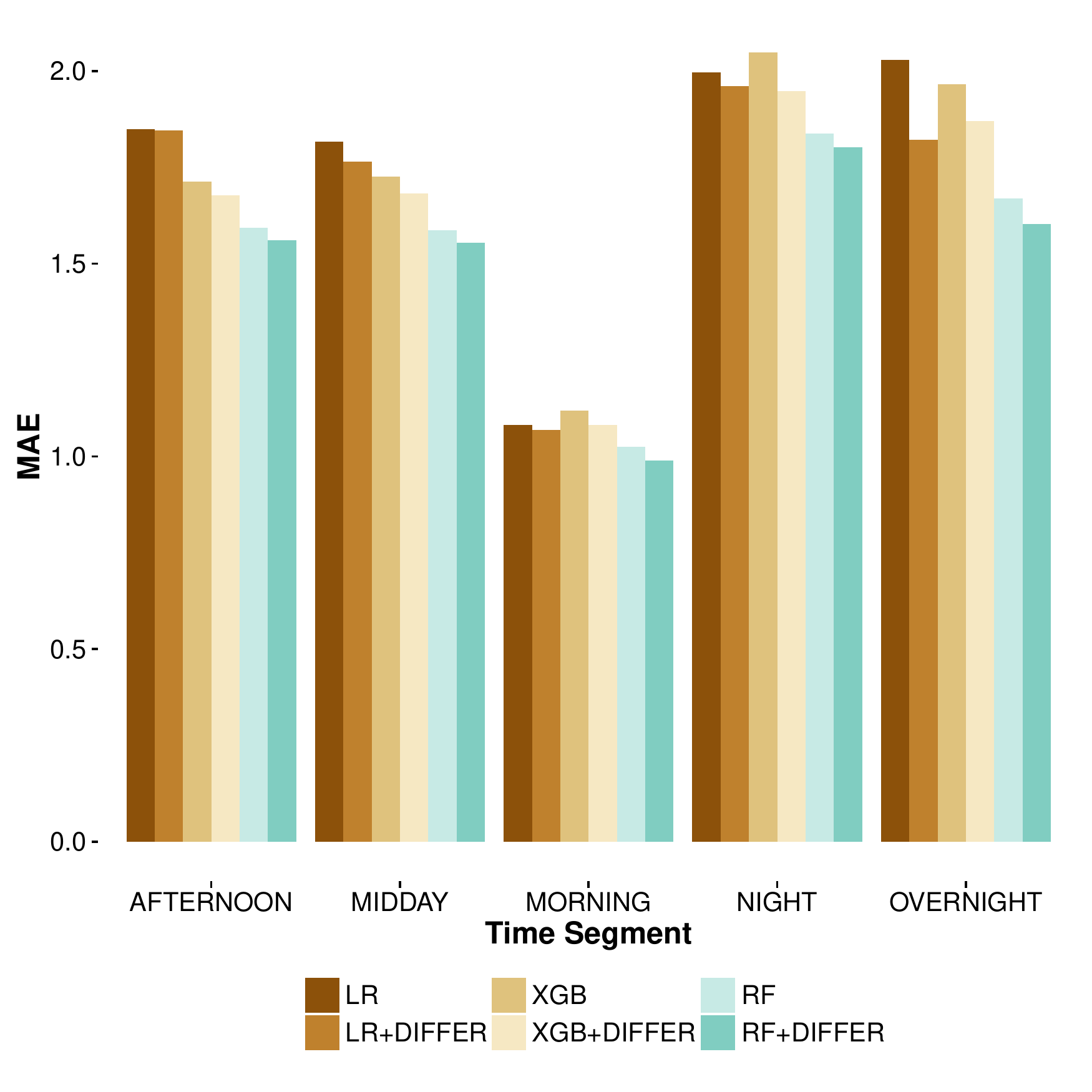}
		\label{fig:2_mae}
		\caption{Chicago}
	\end{subfigure}
\caption{MAE value comparison using different models}
	\label{fig:Features_mae}
\end{figure} 
\begin{figure}[!ht]
	\centering
	\begin{subfigure}[t]{.23\textwidth}
		\centering
		\includegraphics[width=\textwidth]{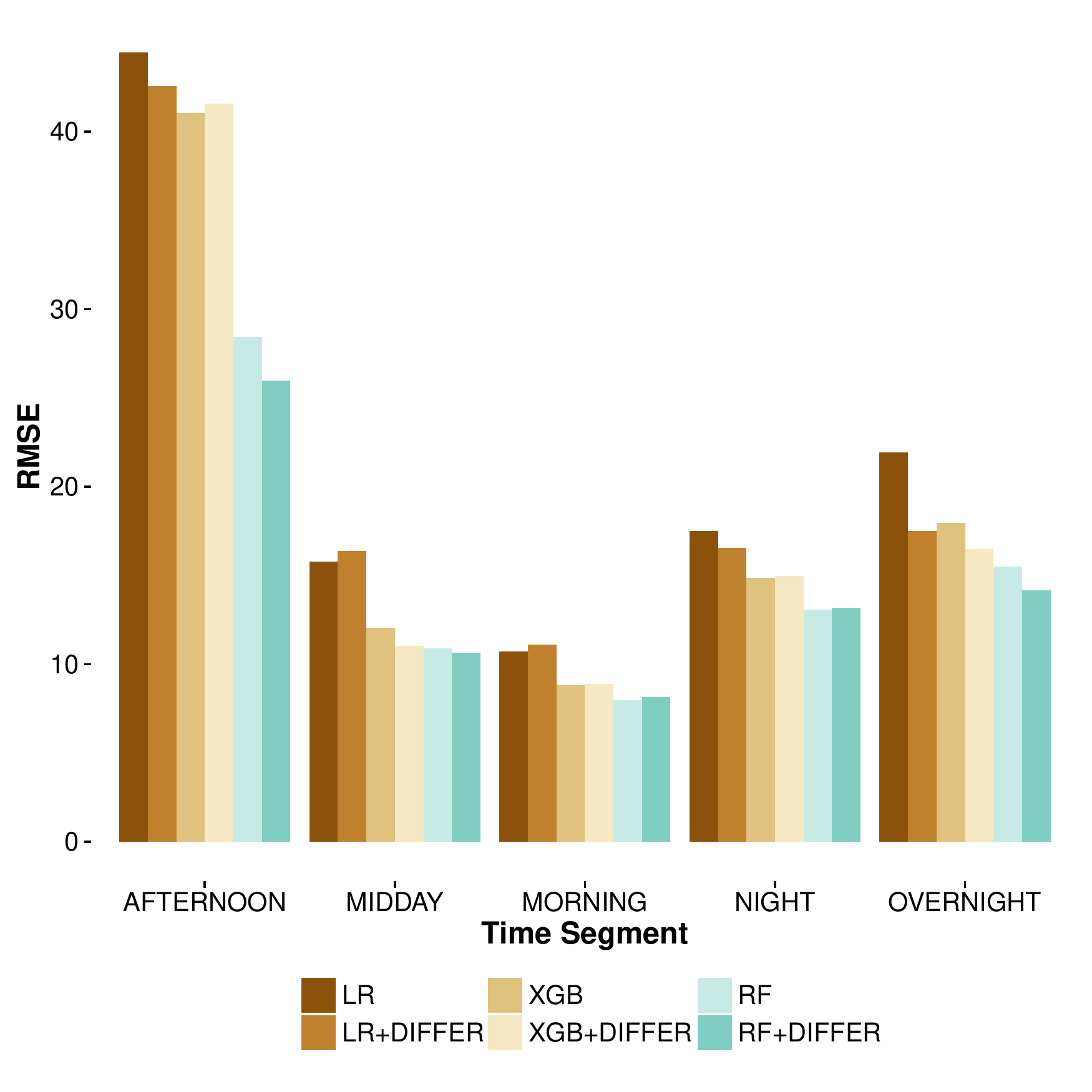}
		\label{fig:1_mae}
		\caption{New York}
	\end{subfigure}
	\centering
	\begin{subfigure}[t]{.23\textwidth}
		\centering
	\includegraphics[width=\textwidth]{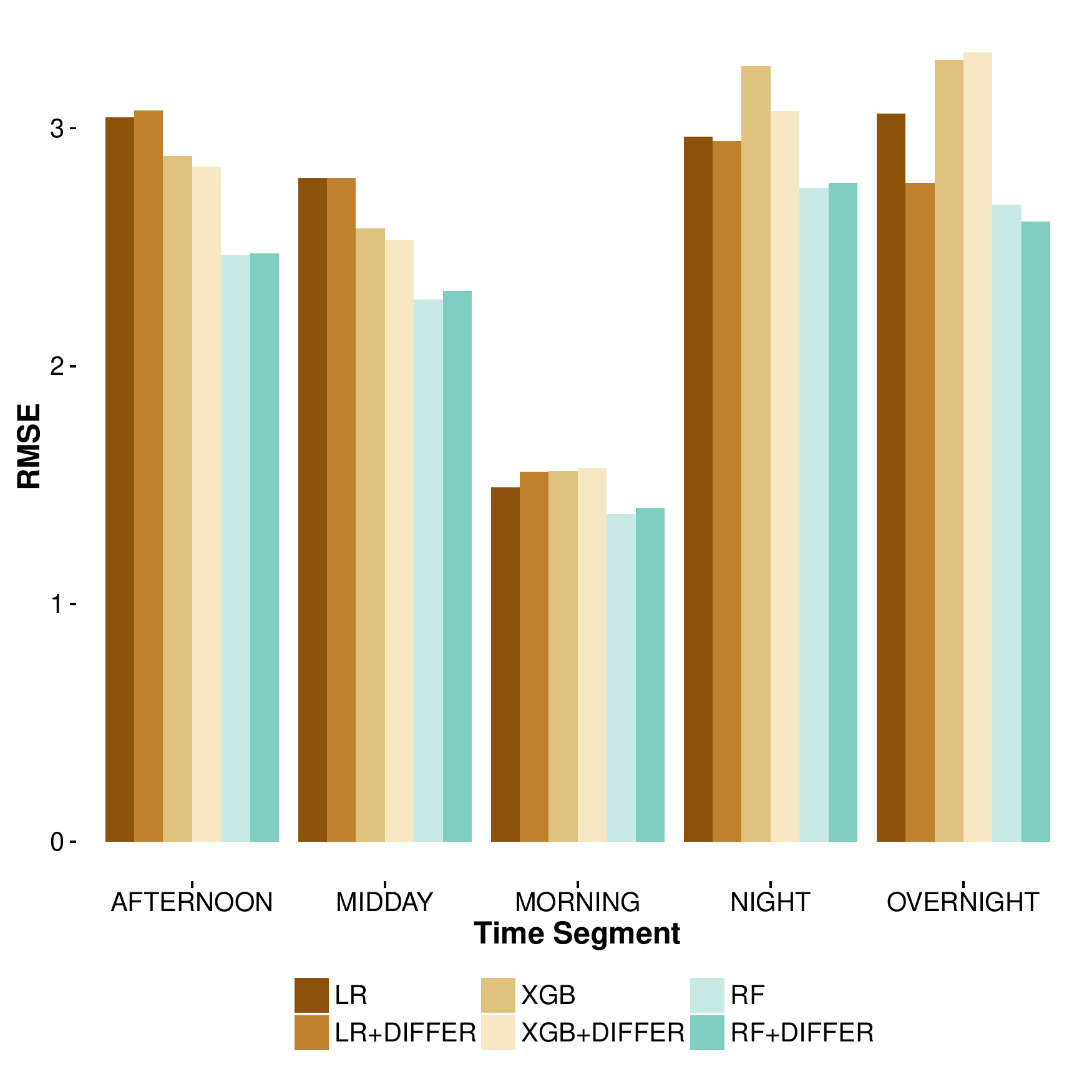}
		\label{fig:2_rmse}
		\caption{Chicago}
	\end{subfigure}
	\caption{RMSE value comparison using different models}
	\label{fig:Features_rmse}
\end{figure} 

\subsection{Performance Study}
 We build two regression models to show the effectiveness of the proposed features, DIFFER, for monthly crime count. The first model is trained with DIFFER and the second model is trained without DIFFER. Finally, we calculate MAE and RMSE value using test set for both models. If the MAE and RMSE is higher for the second model, DIFFER is considered important in monthly crime count. The MAE and RMSE values using different regression algorithms including LR, RF, and XGB are shown in Figures \ref{fig:Features_mae} and \ref{fig:Features_rmse} respectively. The experimental results show that the models with DIFFER have lower errors compared to the models without DIFFER during most of the time intervals. Among three different regression algorithms, RF and XGB show more consistent and better performance. The experimental results verify that the tree-based ensemble learning models are better in predicting crime rate from heterogeneous sources than the linear models. 
\begin{table}[ht]
\caption {Estimated $p$-value for different set of features} 
\label{table:significance_tests}
\begin{subtable}[t]{0.5\textwidth}
\centering
\begin{tabular}{|C{1.2cm}|C{1.2cm}|C{1cm}|C{1.2cm}|C{.8cm}|C{1.1cm}|}
\hline
Time Segment&Historical&Movement&Neighbourhood&POI&DIFFER\\
\hline
Morning &\textbf{3.76e-5} &\textbf{0.003}&$0.968$&$0.827$&\textbf{0.02}\\\hline
Midday&\textbf{4.97e-6}&$0.48$&$0.942$&$0.926$&\textbf{0.008}\\\hline
Afternoon&\textbf{9.79e-4}&$0.45 $&$0.831$&$0.998$&$0.168$\\\hline
Night&\textbf{1.16e-5}&$0.631$&$0.977$&$0.870$&\textbf{0.0403}\\\hline
Overnight& \textbf{5.70e-4}&$0.19$&$0.853$&$0.363$&\textbf{0.002}\\
\hline
\end{tabular}
\caption{New York City}
\end{subtable}\\
\begin{subtable}[t]{0.5\textwidth}
\centering
\begin{tabular}{|C{1.2cm}|C{1.2cm}|C{1cm}|C{1.2cm}|C{.8cm}|C{1.1cm}|}
\hline
Time Segment&Historical&Movement&Neighbourhood&POI&DIFFER\\
\hline
Morning &\textbf{0.007}&\textbf{0.002}&$0.719$&$0.194$&\textbf{1.58e-5}\\\hline
Midday&\textbf{0.0}&\textbf{0.032}&$0.066$&$0.813$&\textbf{6.10e-5}\\\hline
Afternoon&\textbf{0.0}&\textbf{0.009}&$0.318$&$0.772$&\textbf{1.24e-5}\\\hline
Night&\textbf{0.004}&\textbf{0.009}&$0.731$&$0.687$&\textbf{2.34e-4}\\\hline
Overnight&\textbf{0.021}&\textbf{0.006}&$0.550$&$0.242$&\textbf{3.4e-05}\\

\hline
\end{tabular}
\caption{Chicago}
\end{subtable}
\end{table}
\subsection{Significance Test}
As RF shows the best performance result, we measure the significance of each group of features using the RF regression method. First, we train a model with all the features. Next, we train another model without a set of features to identify the effectiveness of that set of features. Finally, we justify the significance of each feature set using a statistical hypothesis test, the paired $t$-test. In this hypothesis test, we set null hypothesis as $MAE_{all}\leq MAE_{wo_i}$. Here, $MAE_{all}$ represents the error with all features and $MAE_{wo_i}$ is the MAE value without a particular feature set $i$. The alternative hypothesis is $MAE_{all}>MAE_{wo_i}$. We gather the sample for t-test using 10-fold cross validation. The estimated $p$-value from the $t$-test is noted in Table~\ref{table:significance_tests} for two cities, New York City and Chicago. When the $p$-value is less than 0.05, it rejects the null hypothesis. We observe that both Historical and DIFFER features have small $p$-value during different time intervals in both cities. It proves that the DIFFER is significant in crime rate prediction.


\section{Conclusion}
This work provides new perspectives to understand crime dynamics with the help of human mobility. It captures the relationship between the monthly aggregated crime data and the movement of people in a region across different periods of the day. The experiments verify that a group of people from a high-risk area will increase the crime risk of their destination.  
\section*{Acknowledgement}
In this paper, we acknowledge the support of the Australian Research Council Discovery \textit{DP190101485}. We also like to acknowledge the help of Phillip Luong in the earlier work of this paper.


\end{document}